\documentclass[breaklinks=true]{aa}

\usepackage{silence}
\usepackage{graphicx}
\usepackage{lscape}
\usepackage{txfonts}
\usepackage{stackengine}
\usepackage{subcaption}
\usepackage{geometry}
\usepackage{float}
\usepackage{tabularx}
\usepackage{longtable}
\usepackage{booktabs}
\usepackage{gensymb}
\usepackage{textgreek}
\usepackage{textcomp}
\usepackage{multirow}
\usepackage{orcidlink}

\usepackage{natbib,twoopt}
\bibpunct{(}{)}{;}{a}{}{,}
\hypersetup{
    colorlinks=true,
    linkcolor=red,
    citecolor=blue, 
    urlcolor=blue 
}

\makeatletter
\let\linenumbers\relax
\let\endlinenumbers\relax
\let\internallinenumbers\relax
\let\resetlinenumber\relax
\makeatother

\begin{document}

   \title{Blazars through the eyes of REDSoX and GOSoX} 
   
   \titlerunning{Blazars through the eyes of REDSoX and GOSoX}
    \authorrunning{S. Capecchiacci et al.}

\author{
  Sara Capecchiacci \inst{\ref*{physics_crete}, \ref*{forth}}\orcidlink{0009-0007-1918-577X} \and
  Ioannis Liodakis \inst{\ref*{forth}}\orcidlink{0000-0001-9200-4006} \and
  Herman L. Marshall \inst{\ref*{MIT}}\orcidlink{0000-0002-6492-1293} \and
  Swati Ravi \inst{\ref*{MIT}}\orcidlink{0000-0002-2381-4184} \and
  Haocheng Zhang \inst{\ref*{baltimore}, \ref*{nasa_goddard}}\orcidlink{0000-0001-9826-1759}
}
\institute{
    Department of Physics, University of Crete, 70013, Heraklion, Greece \label{physics_crete} \and
  Institute of Astrophysics, Foundation for Research and Technology-Hellas, 70013, Heraklion, Greece \label{forth} \and
  MIT Kavli Institute for Astrophysics and Space Research, Massachusetts Institute of Technology, 77 Massachusetts Avenue, Cambridge, MA 02139, USA \label{MIT} \and
  University of Maryland Baltimore County Baltimore, MD 21250, USA \label{baltimore} \and
  NASA Goddard Space Flight Center, Greenbelt, MD 20771, USA \label{nasa_goddard}
}

  \abstract{
    Blazar polarimetric properties provide key constraints on the physical conditions within relativistic jets, offering insights into particle acceleration mechanisms, high-energy radiation processes, and jet composition.
    The Rocket Experiment Demonstration of a Soft X-ray Polarimeter (REDSoX) will serve as a precursor to the Globe Orbiting Soft X-ray Polarimeter (GOSoX), a state-of-the-art polarimeter designed for observations in the soft X-ray range (0.2–0.4 keV), with a planned launch for 2030.
    REDSoX and GOSoX will be the first (and currently only) X-ray polarimeters capable of performing observations below 1 keV, opening a new observational window into high-energy polarimetry.
    In this work, we investigate the soft X-ray polarimetric properties of blazars expected in the REDSoX and GOSoX band and assess the detectability of different samples of sources: one drawn from the RoboPol Monitoring Program, a statistically complete sample of $\gamma$-ray-selected blazars spanning the low-, intermediate-, and high-synchrotron-peaked sources, and one cross-matched between the CAZ (CRTS, ATLAS, and ZTF) and eROSITA catalogs.
    We estimated the expected source fluxes in the selected energy bands through spectral energy distribution fitting performed with the \texttt{Bjet-MCMC} tool.
    We investigated the expected polarimetric properties under three competing emission scenarios: leptonic, hadronic, and hybrid, each one predicting distinct observational signatures.
    We assessed the detectability of every source by comparing our polarization degree predictions with the instrument's minimum detectable polarization function (MDP).
    Finally, we computed the detection probability for each source.
    This work offers a set of promising blazar targets for future observations with GOSoX, highlighting the mission’s potential for probing relativistic jet physics through soft X-ray polarimetry.
}

   \maketitle

\section{Introduction}
\label{sect_introduction}

Blazars are a subclass of active galactic nuclei (AGN) hosting relativistic jets that are oriented close to our line of sight.
Their extreme luminosity is primarily driven by relativistic beaming effects and their emission extends across the entire electromagnetic spectrum.
The observed radiation is thought to originate from particles accelerated to relativistic energies within the jet acceleration region and subsequently cooling as they propagate downstream \citep{blandford2019, hovatta2019, raiteri2025}.
The spectral energy distributions (SEDs) of blazars display a characteristic double-peaked structure, consisting of a low-energy and a high-energy component.
The low-energy peak is associated with synchrotron radiation from relativistic electrons and positrons in the jet, and its characteristic frequency is used to divide blazars into different subclasses. Sources with synchrotron peaks below $10^{14}$ Hz are classified as low-synchrotron-peaked (LSP), those with peaks between $10^{14}-10^{15}$ Hz as intermediate-synchrotron-peaked (ISP), and those with peaks above $10^{15}$ Hz as high-synchrotron-peaked (HSP) blazars \citep{ajello2020fourth}.

The physical origin of the high-energy component of the blazar SED remains an open question.
The main proposed scenarios invoke leptonic and/or hadronic emission processes.
In hadronic scenarios, relativistic protons are expected to play a major role in producing the high-energy emission observed in the X-ray and $\gamma$-ray bands.
The emission can arise from synchrotron radiation of primary protons as well as from secondary charged particles generated through hadronic interactions \citep{mannheim1992,mucke2001}.
Unlike leptonic models, where the two SED components are closely linked, the low- and high-energy emission in hadronic scenarios may originate from different particle populations and therefore do not necessarily show correlated behavior.
In addition, the high-energy component is expected to reach polarization degrees (PD) comparable to (or even exceeding) those of the synchrotron emission at lower energies \cite[e.g.,][]{zhangboettch2013}.
A further implication of hadronic emission models is the production of high-energy neutrinos: relativistic protons interacting with the local radiation field can generate charged pions, whose decay leads to neutrino emission \citep{zhang2019}.

In leptonic models, the high-energy radiation is produced through inverse Compton scattering, where the same population of relativistic electrons responsible for the synchrotron emission upscatters low-energy photons to higher energies. When the seed photons are the synchrotron photons produced within the jet itself, the process is referred to as synchrotron self-Compton (SSC); alternatively, the photons may originate from external radiation fields, such as the broad line region (BLR), molecular clouds, or the dusty torus surrounding the central supermassive black hole, in which case the process is known as external Compton (EC).
Leptonic scenarios generally predict relatively low PD at high energies, since inverse Compton scattering is considered isotropic in blazars, which reduces the polarization of the seed photon population.
As a result, SSC emission is expected to be less polarized than synchrotron radiation, while EC emission is often assumed to be unpolarized because of the thermal origin of the seed photons \citep{bonometto1973,krawczynski2012}.
These scenarios have been favored by a number of studies focused on ISP and LSP sources \cite[e.g.,][]{ehlert2022, middei2022x, peirson2023bllac, marshall2024observations, liodakis2025lsp} and are further supported by the observed correlation between the variability of the synchrotron and high-energy components \citep{rani2013, liodakis2018, liodakis2019, jaeger2023}.
Hadronic models have also been disfavored in individual cases \citep[e.g.,][]{agudo2025}; however, these results cannot yet be generalized to the entire blazar population and a hadronic contribution remains possible in other sources \citep{liodakis2025lsp, Tavecchio2025}, particularly in light of the possible association of very-high-energy neutrino emission with the flaring blazar TXS~0506$+$056 \citep{icecube2018}.

Since different emission scenarios predict substantially different levels of polarization, high-energy polarimetric observations can place significantly stronger constraints on the jet composition than total intensity measurements, which are often unable to distinguish between models \cite[e.g.,][]{boettcher2012modeling}.
Current and future high-energy polarimetry missions will therefore play a key role in advancing our understanding of the physical nature of relativistic jets.

Polarimetric properties of blazars have proved to be of crucial importance in constraining the emission and acceleration mechanisms taking place in these jets, as well as the general jet composition and dynamics.
High-energy observations with the Imaging X-ray Polarimetry Explorer (IXPE, \citealt{weisskopf2022imaging,soffitta2023}) have contributed strongly to the study of HSP sources, especially when matched to simultaneous observations in other bands.
In recent years, a new generation of high-energy polarimeters has been proposed, such as EXPO, e-ASTROGAM \citep{eastrogam2018}, and AMEGO-X \citep{amegox2022}, together with selected or approved missions currently under development, including COSI \citep{tomsick2014}, and eXTP \citep{extp2019}.
Despite these advances, the soft X-ray regime remains largely unexplored from a polarimetric perspective, as no current instrument is capable of performing observations below 1 keV.

The Rocket Experiment Demonstration of a Soft X-ray Polarimeter (REDSoX, \citealt{redsox2018,redsox2023}) and the Globe Orbiting Soft X-ray Polarimeter (GOSoX, \citealt{gosox2021, gosox2026}) have the goal of filling that gap, providing unprecedented spectropolarimetric observations in the soft X-ray band, between $0.2$ and $0.4 \ \rm keV$.

In this work, we aim to indicate a new sample of potential blazar targets for the GOSoX mission, following the detectability framework built in \citealt{cape2026}.
Starting from the sample of blazars in the RoboPol monitoring program \citep{blinov2019}, the BlazEr1 eROSITA blazar catalog \citep{erosita2026} and the CAZ catalog \citep{kouch2025}, we considered three different proposed emission scenarios to estimate the expected PD of the sources: leptonic, hadronic, and a hybrid one between the two.
We compared our predictions with the minimum detectable polarization function (MDP) of REDSoX to assess their detectability in the $0.2-0.4 \ \rm keV$ band.
GOSoX will follow REDSoX with a very similar level of performance; therefore, our study serves as a benchmark for this mission.

In Section \ref{sect_gosox} we describe the instrument and its main science goals.
In Section \ref{sect_sources} we describe our sample and data.
In Section \ref{sect_data_analysis} we describe the analysis performed on the sample.
We present our results in Section \ref{sect_results}, along with a summary and discussion in Section \ref{sect_conclusions}.

\section{REDSoX and GOSoX}
\label{sect_gosox}

REDSoX is a NASA sounding rocket mission selected in 2022 for launch in 2027.
It will serve as a technology demonstration and pathfinder mission for its orbital successor GOSoX, a NASA Pioneer-class mission selected in 2026 for launch in 2030.
GOSoX will perform soft X-ray polarimetric studies of numerous astrophysical objects, such as blazars, quasars, X-ray binaries, transient X-ray binaries (XRBs), neutron stars, and accreting white dwarf binaries.
While REDSoX is primarily a technology demonstrator, its instrumental characteristics are expected to be broadly representative of those of GOSoX.
REDSoX will observe Mrk~421 during a 30-minute parabolic sounding rocket flight with approximately 5 minutes of science observations during its flight apex.
Our study uses the REDSoX MDP to provide a sample of blazar targets for GOSoX, which will have similar instrumental properties, but with exposures of $0.1-3 \ \rm Ms$.

Both REDSoX and GOSoX employ full-shell nested Wolter-I grazing-incidence mirrors to collect and focus soft X-rays.
They will utilize critical-angle transmission (CAT gratings; \citealt{garner2021,heilmann2025,ravi2026}) to perform spectroscopy.
The mirror assembly focuses light into three distinct polarimetry channels rotated by $120^\circ$ to collect full linear Stokes information.
Each polarimetry channel consists of an assembly of transmission gratings and laterally graded multilayer-coated mirrors (LGMLs; \citealt{marshall2014}), which act as Bragg reflectors at the Brewster angle.
The multilayered coatings increase the reflectivity, while their lateral grading extends the bandpass by allowing the dispersed light to reflect from the region of the mirror whose multilayer spacing is optimized for the corresponding wavelength.
At the Brewster angle ($\approx 45^\circ$ for X-rays), the reflected light is fully linearly polarized, with the electric vector perpendicular to the reflection plane.
Each multilayer coated flat mirror focuses on a CCD detector to measure the intensity reflected from that mirror.
Orienting three flats at position angles $120 ^\circ$ apart, the instrument will measure the three Stokes parameters $I$, $Q$, $U$ without the need for rotations.
The resulting instrument has a spectral resolving power of $E / \delta E > 100$, corresponding to the ability to distinguish spectral features separated by less than about 1\%.
The X-ray full-shell focusing optics are being manufactured at NASA MSFC \citep{bongiorno2025}, gratings at MIT SNL, and LGMLs at Lawrence Berkeley National Laboratory.
The instrument components are currently undergoing testing at the MIT Polarimetry Beamline \citep{heine2024,heine2025,heine2026} and will eventually undergo end-to-end testing at the NASA MSFC 100 m beamline.
GOSoX will follow the same principle, but it will have a reduced focal length (from $2.5 \ \rm m$ to $1.2 \ \rm m$).

\section{Samples}
\label{sect_sources}

We selected our first blazar sample from the RoboPol monitoring program \citep{blinov2019}.
The sources were monitored in the R band with the RoboPol polarimeter \citep{robopol} mounted on the 1.3 m telescope at the Skinakas Observatory in Crete, Greece.
Between 2013 and 2017, the program obtained measurements of the PD and polarization angle (PA) for a total of 222 AGN \citep{blinov2016, blinov2018}, with the primary goal of investigating possible connections between rotations of the PA and $\gamma$-ray flares.
From this sample, we selected the statistically complete subsample of gamma-ray-loud blazars, resulting in a total of 62 sources: 36 LSP, 13 ISP, and 13 HSP.
In addition, we included two extra sources outside the RoboPol Monitoring Program that had previously been identified as promising targets for GOSoX observations, namely Mrk~421 and 3C~273.

A second sample of sources was obtained by cross-matching the CAZ \citep{kouch2025} and the eROSITA BlazEr1 \citep{erosita2026} catalogs and selecting sources with fluxes higher than $10^{-12} \ \rm  erg \ cm^{-2} \ s$ in the $0.2-0.5 \ \rm keV$ band (reported in BlazEr1).
This selection led us to a sample of 63 sources: 6 LSP, 6 ISP, and 51 HSP.

\section{Data analysis}
\label{sect_data_analysis}

\subsection{Sources from the RoboPol monitoring program}
\label{sect_rbpl_sources}

To derive flux and polarization predictions for the sources of the RoboPol monitoring program, we followed the framework described in detail in \citealt{cape2026}.
Here, we provide a brief summary of the methodology followed.

The multiwavelength flux data were collected from the Markarian Multiwavelength Data Center (MMDC) and the Space Science Data Center (SSDC) SED Builder tool.
We began by fitting the frequency-binned data using the \texttt{Bjet-MCMC} framework \citep{bjetmcmc}.
For two HSP sources, namely Mrk~421 and 1ES~1959+650, we were unable to obtain accurate fits with the \texttt{Bjet-MCMC} tool.
Therefore, we fitted with the JetSeT tool \citep{tramacere2009,tramacere2011,tramacere2020}, which provided more accurate results in these specific cases.
The fitted SED was then shifted for the synchrotron peak to match the value reported in the Fermi 4LAC-DR3 \citep{ajello2020fourth} and for the flux in the $0.3-10 \ \rm keV$ range to match the value reported in the CAZ catalog \citep{kouch2025}.

\begin{figure}[h!]
    \centering
    \includegraphics[width=\linewidth]{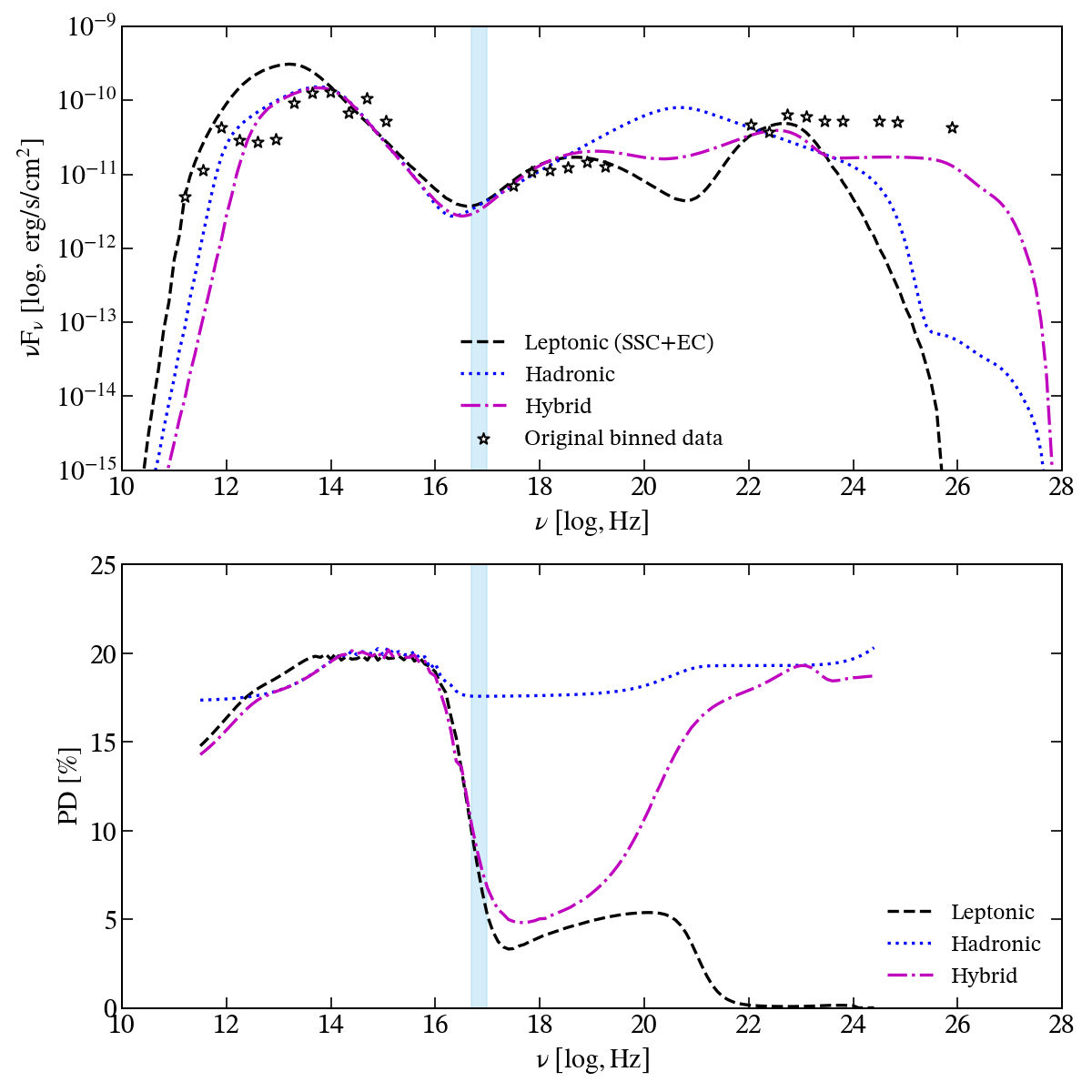}
    \caption{
    The BL~Lac models used as a template for all our LSP sources.
    \textit{Top:} Fitted SED. Black stars represent the frequency-binned data. \textit{Bottom:} Predicted PD.
    The dotted blue line represents the hadronic model, the dashed black line represents the leptonic, and the dash-dotted magenta line the hybrid.
    The blue shaded area highlights the GOSoX observation band (0.2-0.4 keV).
    }
    \label{fig:bllac_model}
\end{figure}

Following \citealt{cape2026}, we took into account three distinct polarization models: leptonic, hadronic, and a hybrid one between the two.
In all three cases, the low-energy spectral component is assumed to originate from electron synchrotron emission, while the high-energy component differs depending on the emission scenario considered.
The leptonic scenario includes polarized SSC emission together with an unpolarized EC component.
The hadronic scenario accounts for synchrotron emission from both primary protons and secondary electron–positron pairs, while the hybrid scenario combines the hadronic emission with an additional SSC contribution.
The polarization models are based on fits to observations of three representative sources: BL~Lac (LSP), BZB~J0211+1051 (ISP), and Mrk~501 (HSP).
These sources were used as templates for each SED class, and their polarization models were subsequently shifted to reproduce the observational properties of the targets in our sample.
Fig. \ref{fig:bllac_model} represents the models obtained for BL~Lac; the other two templates are shown in Appendix \ref{appendix_polarisation_models}.

To obtain polarization models compatible with current observations, we applied scaling factors in both frequency and PD to the template.
The frequency scaling was determined from the position of the minimum between the two bumps of the SED, while the polarization scaling was chosen so that the model reproduces the median R-band PD measured by RoboPol during the monitoring campaign. 

We calculated the MDP of GOSoX following the procedure described in Sect. \ref{sect_mdp} and compared it with our flux and PD predictions to assess the detectability of each source.
Results are reported in Fig. \ref{fig:results}.
Figure \ref{fig:cosi_comparison} compares the PD predicted at the central energies of the GOSoX and COSI bands, highlighting the potential of simultaneous polarimetric observations to discriminate between competing emission models.
We refer the reader to \citealt{cape2026} for a more detailed description of the methodology.

\subsection{Sources from CAZ and BlazEr1 catalogs}
\label{sect_sources_caz_erosita}

For the sources in the CAZ and BlazEr1 catalogs, we cross-matched the two, obtaining a sample of 63 sources (6 LSP, 6 ISP, and 51 HSP), none of which overlap with the RoboPol sample.
We note that the peak frequencies reported in CAZ match the ones reported in 4LAC-DR3.

For the SED of these sources, we used the SED of the three template blazars reported in Sect. \ref{sect_rbpl_sources} as reference for all of the sources in the same SED class (i.e., the Mrk~501 SED was used for all HSP, etc.), applying horizontal and vertical shifts to match the flux at 5.0 keV (the central energy of the energy band reported in CAZ) and the synchrotron peak frequency to the values reported in CAZ.

\begin{figure}[h!]
    \centering
    \includegraphics[width=0.9\linewidth]{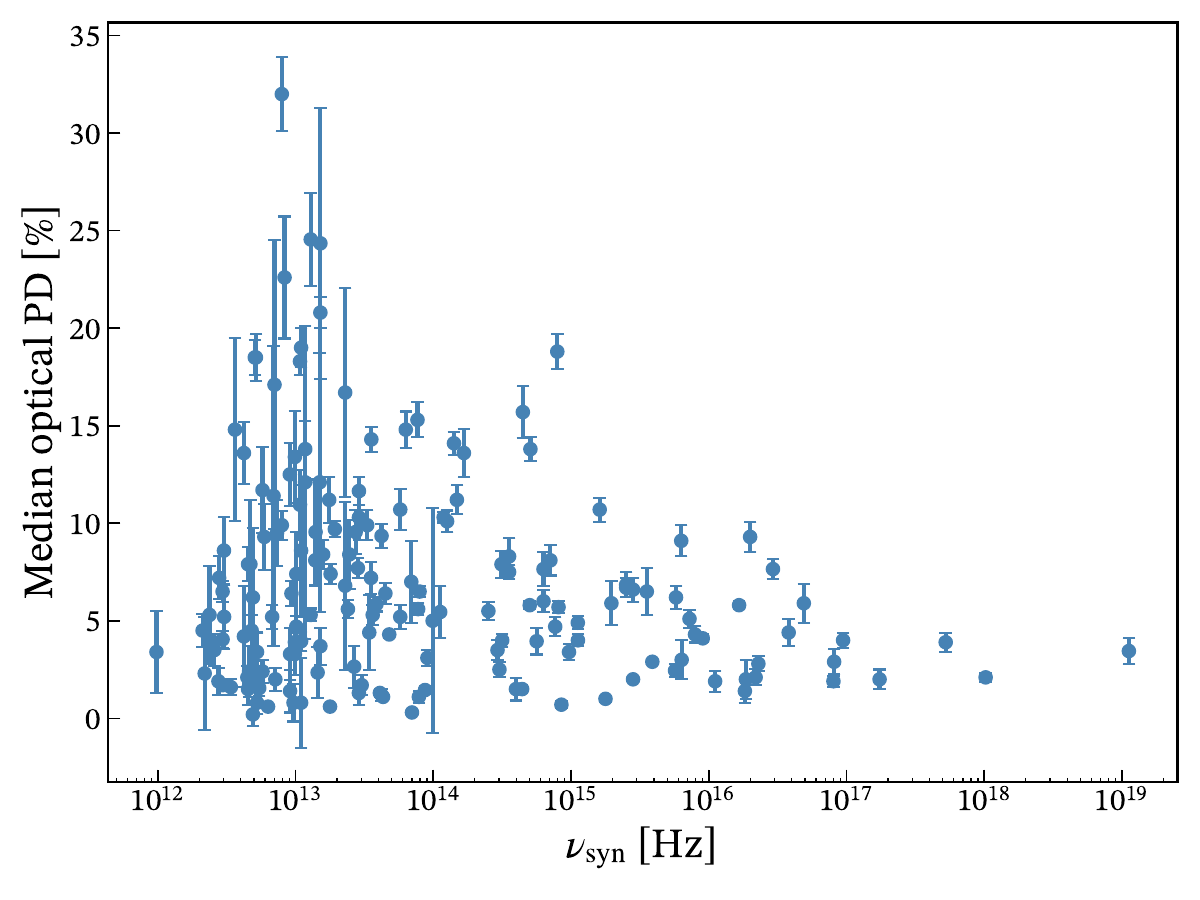}
    \caption{Median optical PD of each source in the RoboPol sample, calculated from all available PD measurements, plotted as a function of the synchrotron peak frequency reported in 4FGL-DR3.
    }
    \label{fig:pd_vs_peak}
\end{figure}

\begin{figure}[h!]
    \centering
    \includegraphics[width=0.9\linewidth]{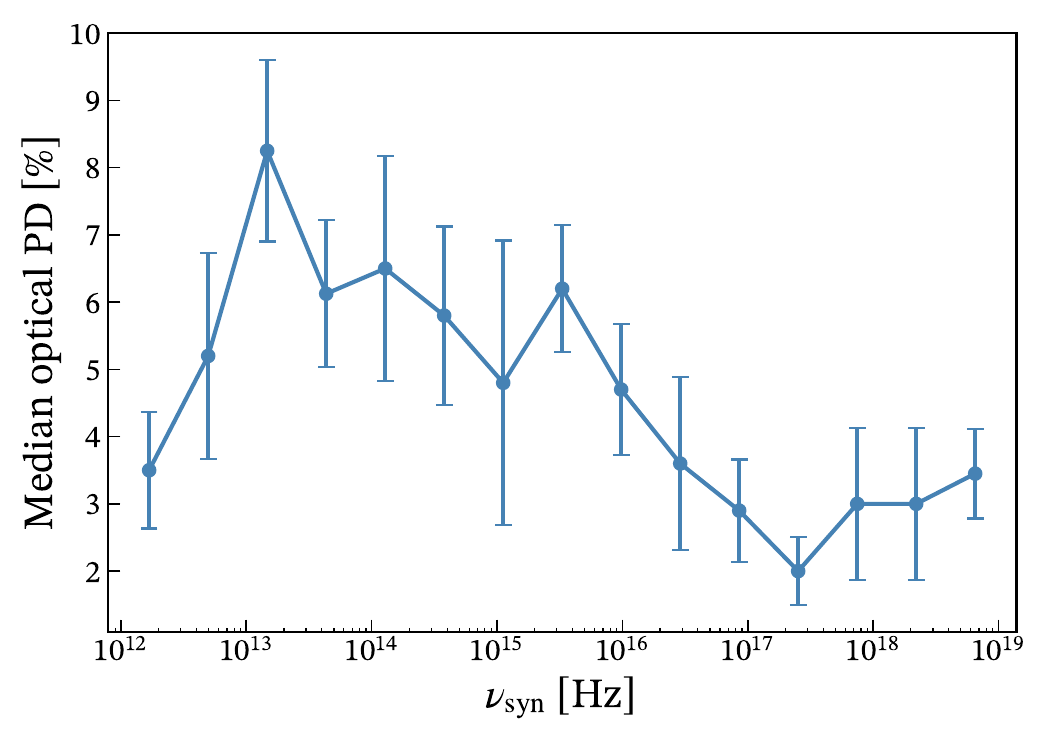}
    \caption{Median optical PD of the RoboPol sources, calculated from all available PD measurements for each source and subsequently binned by synchrotron peak frequency, plotted as a function of the synchrotron peak frequency reported in 4FGL-DR3.
    }
    \label{fig:pd_vs_peak_binned}
\end{figure}

For the PD models, these sources were not part of the RoboPol monitoring program and therefore lacked the optical PD measurements required to shift the templates.
To estimate their optical PD, we considered all sources in the RoboPol monitoring sample with synchrotron peak frequencies reported in the 4LAC-DR3 catalog and investigated the relationship between the median optical PD and synchrotron peak frequency (Fig. \ref{fig:pd_vs_peak}).
We then binned the sources by peak frequency and computed the median PD and its uncertainty in each bin.
The resulting trend (Fig. \ref{fig:pd_vs_peak_binned}) was used to assign an optical PD to each source based on its synchrotron peak frequency.
Using these estimated optical PD values, we shifted the PD templates following the procedure described in Sect. \ref{sect_rbpl_sources} and obtained GOSoX-band PD predictions for the three emission models.

For the 51 HSP selected with these criteria, we also adopted an alternative approach to predict the X-ray polarization degree.
From the analysis of the IXPE HSP sample presented in \cite{cape2025}, we found that the ratio between the X-ray and optical PD has a median value of 2.5.
After assigning an optical PD to each HSP using the relation shown in Fig. \ref{fig:pd_vs_peak_binned}, we estimated the X-ray PD by scaling the optical value by this factor.
Although this prediction is independent of the adopted polarization model, it could provide a more realistic estimate, as it is empirically based on previous X-ray and optical polarization observations of HSP blazars.

The predictions resulting from both approaches are reported in Appendix \ref{appendix_results_caz_erosita}.

\subsection{Minimum detectable polarization}
\label{sect_mdp}

The GOSoX minimum detectable polarization (MDP) is shown in Fig.~\ref{fig:results}, assuming an exposure time of 1Ms for LSPs and ISPs (top and middle panels, respectively) and 300ks for HSPs (lower panel).
Variations in the PA during an observation can introduce depolarization, reducing the measured PD and potentially leading to nondetections.
Following \cite{cape2026}, we assume that the PA remains approximately stable over timescales of 14 days for all three SED classes, and up to 28 days in some cases.
While faster PA variations and rotations can occur, this assumption provides a reasonable simplification for estimating source detectability.
The exposure times adopted here fall within these timescales, limiting the expected impact of PA variability under this assumption.

The MDP curves were computed in the $0.2-0.4 \ \rm keV$ band using the energy-dependent effective area provided by the instrument auxiliary response file (ARF).
For a grid of integrated energy fluxes between $10^{-15}$ and $10^{-7} \ \rm erg \ cm^{-2} \ s^{-1}$, we assumed a flat energy-flux density across the considered energy band.
The corresponding photon flux density was obtained by dividing the energy-flux density by the photon energy.
The total source count rate was then calculated as 
\begin{equation}
R_{\rm S} = \sum A_{\rm eff}(E)\, n_E(E)\, \Delta E
\end{equation}
where $A_{\rm eff}(E)$ is the effective area, $n_E(E)$ is the photon flux density, and $\Delta E$ is the energy-bin width.

The complete REDSoX/GOSoX instrument model \citep{ravi_2026_mdpc} includes both an energy-dependent modulation response $\mu(E)$ and a background contribution. The MDP at 99\% confidence is written as \begin{equation}
    {\rm MDP}_{99}(E)[\%] =
    \frac{4.291}{\mu(E) R_{\rm S}}
    \sqrt{\frac{R_{\rm S}+R_{\rm B}}{T}}\times 100,
\end{equation}
where $R_{\rm S}$ and $R_{\rm B}$ are the source and background count rates, respectively, $T$ is the exposure time in seconds, and $\mu(E)$ is the energy-dependent modulation factor. The energy-dependent modulation factor has modeled values of approximately $\mu\simeq0.92$--$0.96$ across the $0.2$--$0.4$~keV band and may therefore be approximated with a constant representative value of $\mu=0.95$.

For the sensitivity estimates presented here, we neglect the background contribution and approximate the wavelength-dependent modulation response with a constant $\mu=0.95$. Under these assumptions, the expression can be reduced to
\begin{equation}
    {\rm MDP}_{99}[\%] =
    \frac{4.291}{\mu\sqrt{R_{\rm S}T}}\times 100.
\end{equation}
The resulting curves provide an estimate of the expected polarization sensitivity as a function of the integrated
$0.2$--$0.4$~keV source flux for the adopted exposure times.

\subsection{Detection probability}
\label{sect_duty_cycle}

For each of the sources considered, we computed the instrument duty cycle, defined as the probability of detecting the source in a blind survey.
This quantity provides a useful benchmark for prioritizing targets according to the expected detection probability.
We note that in this framework, the uncertainties on the obtained PD and flux predictions strictly depend on the adopted catalogs used for SED and polarization model shifting.

For the RoboPol monitored sources, we estimated the duty cycle through a Monte Carlo approach.
For each source, 1000 X-ray flux realizations were generated by drawing from a Gaussian distribution centered on the $0.3-10 \ \rm keV$ flux reported in the CAZ catalog, with standard deviation equal to the corresponding flux uncertainty.
The simulated CAZ fluxes were then converted to the REDSoX energy band using the conversion factor derived from the source SED.
Nonpositive flux realizations were rejected and redrawn.
Independently, 1000 optical PD realizations were obtained by randomly sampling the observed RoboPol polarization measurements of the source.
The corresponding X-ray PD was estimated by applying the model-predicted optical-to-X-ray scaling.
Finally, each simulated flux--PD pair was compared with the REDSoX MDP, and the duty cycle was defined as the fraction of realizations for which the predicted X-ray polarization exceeded the MDP.

For the unmonitored sources, which lack optical polarization measurements, we generated 1000 optical PD realizations by drawing from a Gaussian distribution centered on the median polarization degree of the corresponding synchrotron-peak-frequency bin shown in Fig.~\ref{fig:pd_vs_peak_binned}, with a standard deviation that is equal to the uncertainty on the median within that bin.
The optical PD was then rescaled to the GOSoX band as described in Sect. \ref{sect_sources_caz_erosita}.
The X-ray flux realizations were generated following the same procedure adopted for the monitored sources.
The duty cycle was then computed following the same procedure described above.

\section{Results}
\label{sect_results}

\subsection{Promising candidates}

\begin{figure}[h!]
    \centering
    \begin{subfigure}{0.9\linewidth}
        \centering
        \includegraphics[width=\linewidth]{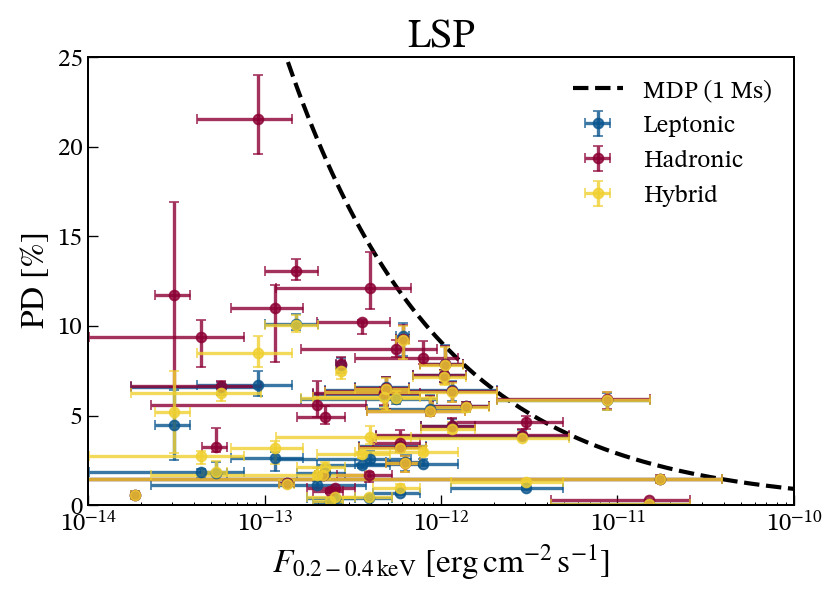}
        \label{fig:lsp_mdp}
    \end{subfigure}
    \hfill
    \begin{subfigure}{0.9\linewidth}
        \centering
        \includegraphics[width=\linewidth]{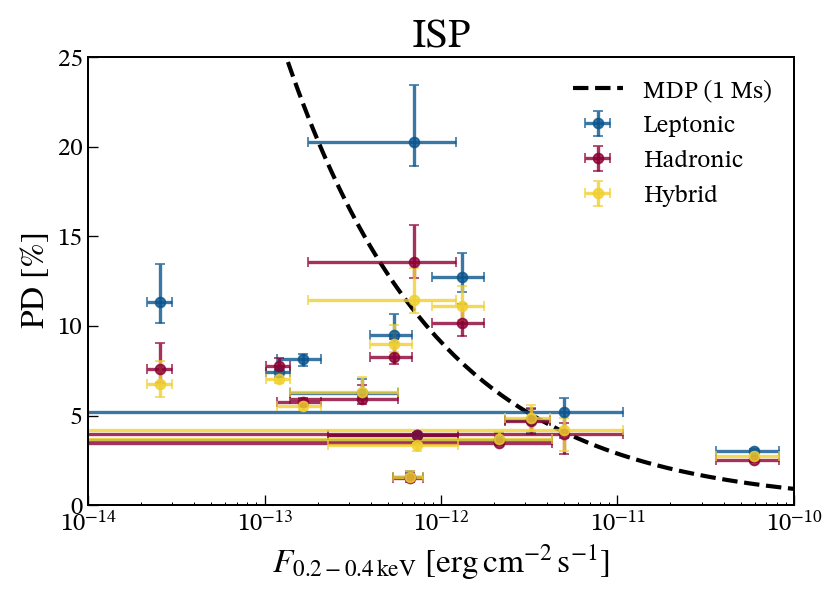}
        \label{fig:isp_mdp}
    \end{subfigure}
    \hfill
    \begin{subfigure}{0.9\linewidth}
        \centering
        \includegraphics[width=\linewidth]{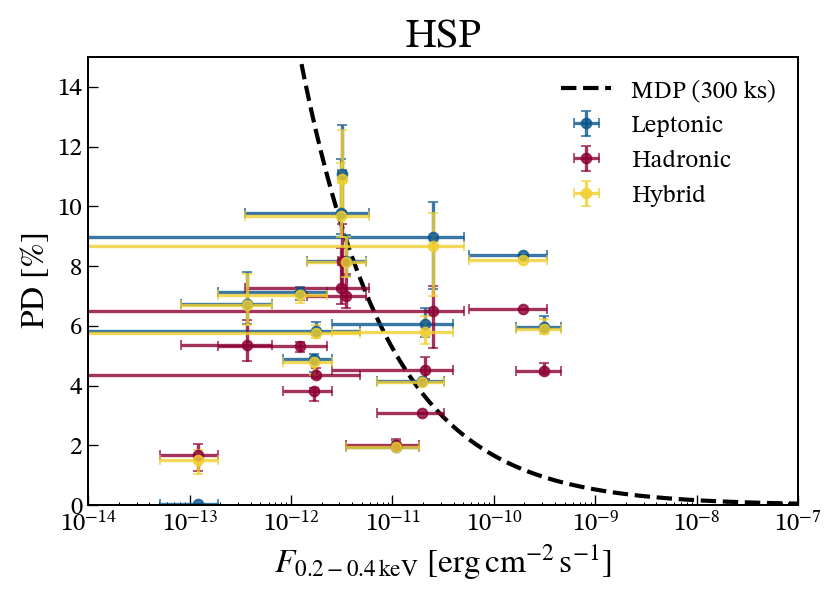}
        \label{fig:hsp_mdp}
    \end{subfigure}

    \caption{
    Results of our predictions for the RoboPol monitored sources (top to bottom: LSP, ISP, HSP).
    Red dots represent the hadronic model, blue dots the leptonic model, and yellow dots the hybrid model.
    The black dashed curve represents the MDP calculated for a given exposure time (300 ks for HSP, 1 Ms for LSP and ISP).
    }
    \label{fig:results}
\end{figure}

In Fig. \ref{fig:results}, we report the results obtained for the RoboPol monitored sources; the tables listing the sources with the highest duty cycle are given in Appendix \ref{appendix_results_robopol}.
The results for the CAZ--BlazEr1 matched sources can be found in Appendix \ref{appendix_results_caz_erosita}.
The complete tables are available on the \href{https://zenodo.org/records/23057566}{Zenodo} platform.
We note that for ISP and LSP sources, for which the REDSoX band falls into the synchrotron bump of the SED, we only report on results obtained from the leptonic model; for the HSP, where the observation band falls into the high-energy bump, we report all three models.
For HSP sources, generally brighter in the soft X-ray band, we chose an exposure time of 300 ks ($\approx$3 days), due to their high variability and high fluxes in this energy band.
For ISP and LSP sources, we choose 1 Ms ($\approx$11 days).
LSP sources are expected to be hard to detect in the $0.2-0.4 \ \rm keV$ band, due to their low brightness in the X-ray range.
On the other hand, ISP and HSP sources appear to be better candidates for GOSoX observations.
This is particularly true for HSP blazars, as the GOSoX energy band probes the synchrotron component near its peak, where higher polarization degrees are expected.
Amongst the RoboPol HSP sources, the most promising candidates are Mrk~421 and 1ES~1959+650, both of which have previously been observed by IXPE; amongst the LSP sources, BL~Lac is the one with the highest duty cycle; for the ISP, TXS~2147$-$144 and S4~1749$+$70 have a high chance of being detected.
Amongst the CAZ-eROSITA matched blazars, many HSP sources present extremely high duty cycles ($>90 \%$), the highest being that of 4FGL~J1057.8$-$2754.
This is also the case for LSP sources, where the source with the highest duty cycle is 4FGL~J0433.0$+$0522.
As for the ISP, 4FGL~J0847.2$+$1134 and 4FGL~J1213.3$-$2618 are characterized by the highest duty cycle.

\subsection{Multiwavelength polarimetric synergies}
\label{sect_ixpe_comparison}

The GOSoX and IXPE energy bands probe two particularly informative regions of the blazar spectral energy distribution, especially in the case of HSP blazars.
In these sources, the GOSoX band samples the high-energy end of the synchrotron component, while the IXPE band probes the low-energy end of the high-energy emission component.
As shown by \cite{peirson2022}, comparing the polarization properties measured in these two bands provides a powerful diagnostic of the jet composition.
This is crucial to discriminate between hadronic and leptonic models, particularly in the case of LSP and ISP sources, where the predicted PD in the two bands differs significantly based on the considered model \citep{cape2026}.

\begin{figure}[h!]
    \centering
    \begin{subfigure}{0.9\linewidth}
        \centering
        \includegraphics[width=\linewidth]{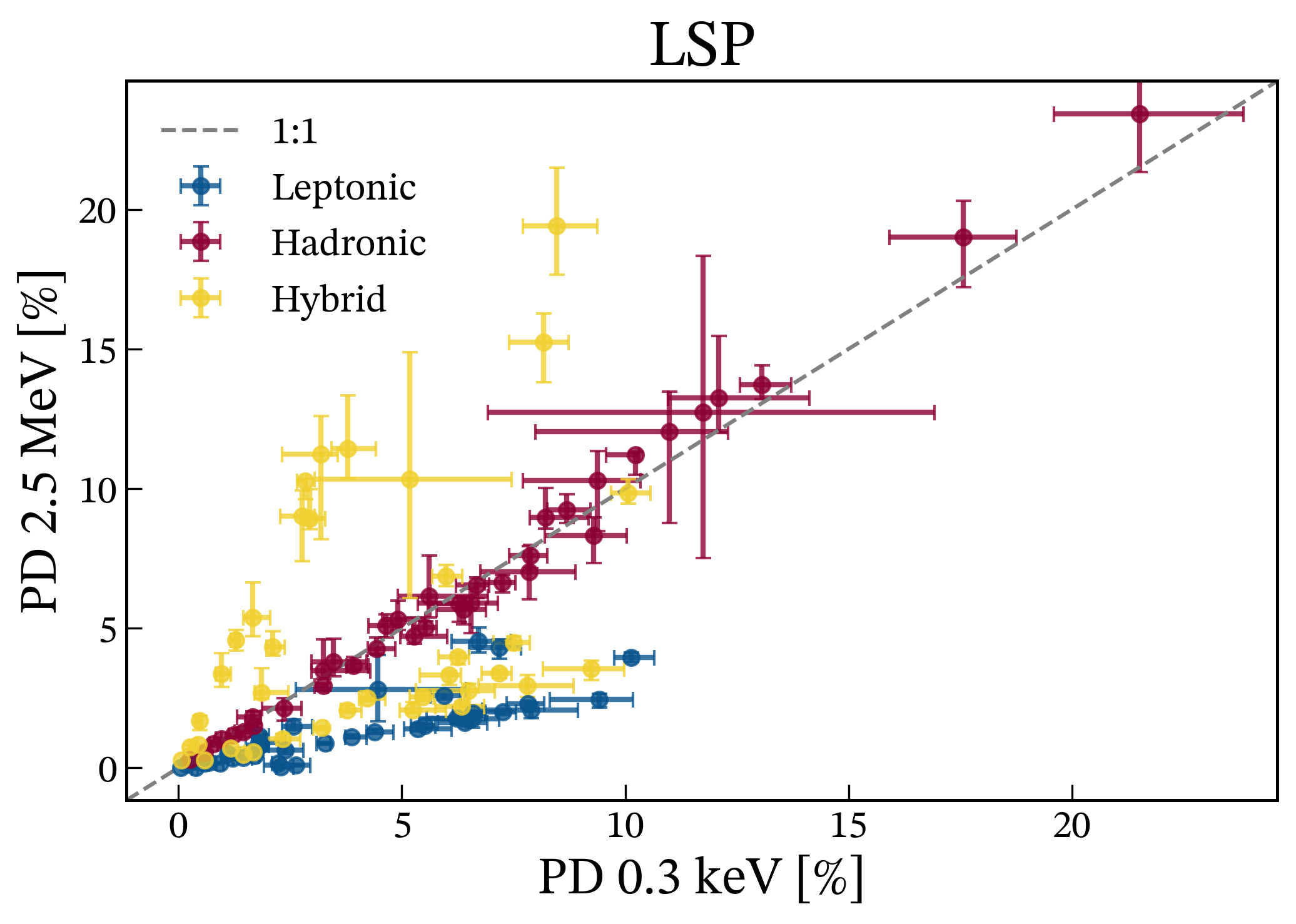}
    \end{subfigure}
    \hfill
    \begin{subfigure}{0.9\linewidth}
        \centering
        \includegraphics[width=\linewidth]{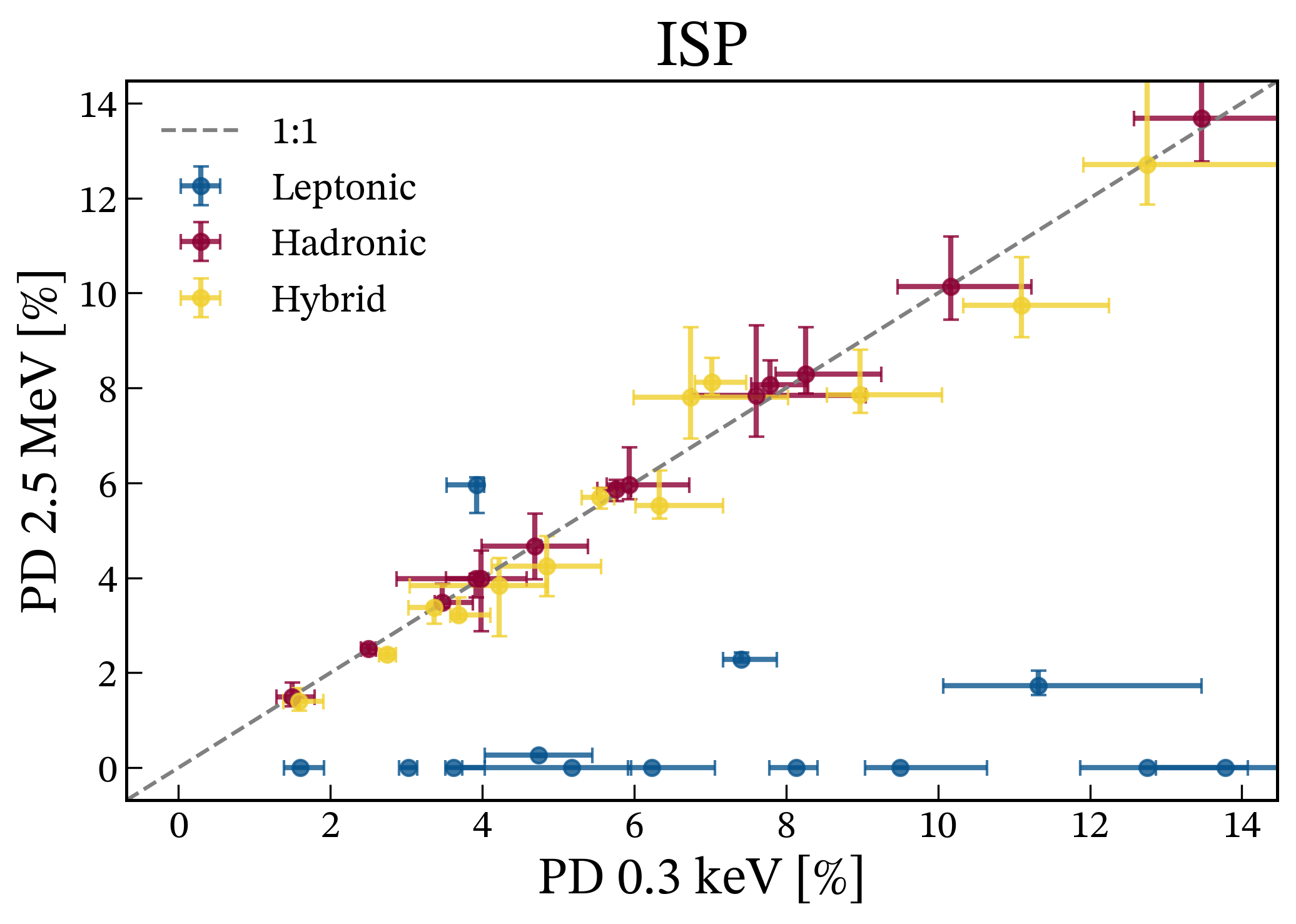}
    \end{subfigure}
    \hfill
    \begin{subfigure}{0.9\linewidth}
        \centering
        \includegraphics[width=\linewidth]{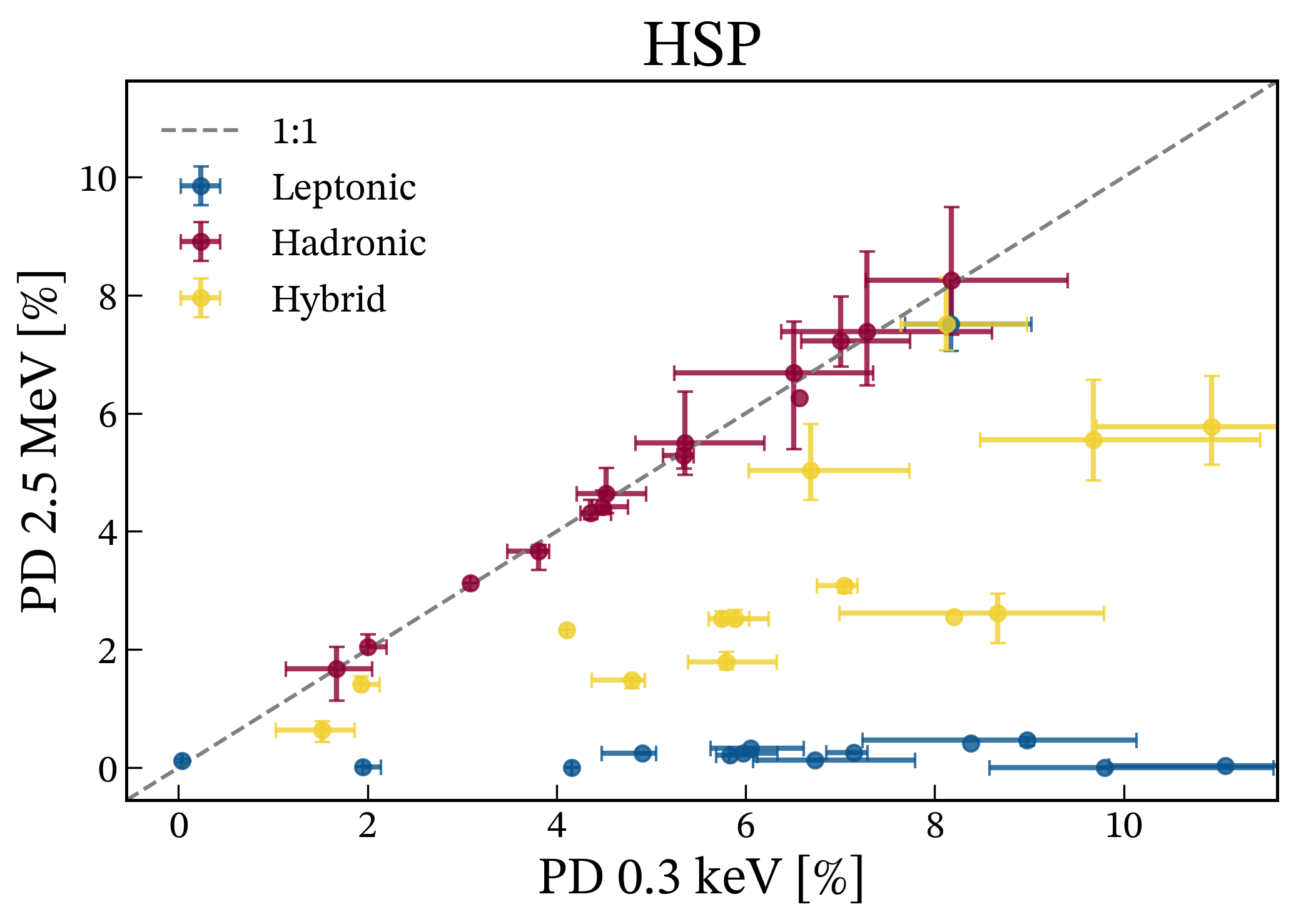}
    \end{subfigure}
    \caption{
    Comparison between the predicted PD in the GOSoX band ($x$ axis) and the COSI band ($y$ axis) for LSP, ISP and HSP sources (top to bottom) of the RoboPol sample.
    Blue points represent the leptonic model, red points the hadronic, and yellow the hybrid.
    The dashed line represents equality between the two predictions.
    }
    \label{fig:cosi_comparison}
\end{figure}

Besides IXPE, several polarimetric missions covering complementary energy bands are expected to operate during the GOSoX era, providing opportunities for simultaneous multiwavelength polarimetric observations.
In particular, COSI \citep{tomsick2014}, with launch planned for 2027, will provide polarimetric measurements in the $0.2-5 \ \rm MeV$ band.
Fig.~\ref{fig:cosi_comparison} reports the comparison between the PD predictions in the GOSoX and the COSI bands.
A more detailed discussion of COSI's capabilities for blazar observations can be found in \citet{cape2026}.

For LSP sources, the hadronic model generally reaches out to higher PD results for both bands, but these sources already proved to be the faintest in the soft X-ray band.
Therefore, they are not considered to be optimal candidates for observations with GOSoX.
For some HSP, the PD predicted by the leptonic model is close to 0\% in the COSI band, but higher in the GOSoX band (reaching up to $\approx 11 \%$).
Therefore, observing these sources simultaneously in the $0.2-0.4 \ \rm keV$ and $0.2-5 \ \rm MeV$ bands could prove crucial to discriminate between a leptonic scenario and the proposed alternatives.
A similar reasoning can be applied to the ISP sources, where there is a very clear distinction between the leptonic model and the other two.

\section{Conclusions}
\label{sect_conclusions}

In this work, we applied the framework from \citealt{cape2026} to obtain predictions on expected flux and PD in the GOSoX and REDSoX band ($0.2-0.4 \ \rm keV$) for two independent, nonoverlapping samples of blazars: one drawn from the RoboPol Monitoring Program and the other obtained by cross-matching the CAZ and BlazEr1 catalogs.
We offered PD predictions under three different polarization scenarios: leptonic, hadronic, and hybrid.
We computed the MDP of REDSoX, which will act as a precursor for GOSoX and is expected to have very similar performance capabilities, by taking into account its effective area and different exposure times for different SED classes.
The exposure times adopted in this work were $300 \ \rm ks$ for HSP blazars, which are typically brighter and more variable in the X-ray band, and $1 \ \rm Ms$ for ISP and LSP with the aim to increase their detection probability.
These values were chosen as representative observing times and can be easily modified within the code to obtain updated detectability predictions for different observing strategies.

We assessed the detectability of our source sample by comparing the predicted X-ray PD with the instrument's MDP.
This analysis identifies a set of promising candidates for GOSoX observations, supporting the optimization of future observing strategies.
We also computed the duty cycle of each source, defined as the probability of detection in a blind survey, which provides a quantitative metric for prioritizing future targets.
These predictions are subject to several limitations arising from the assumptions adopted in our methodology. In particular, the use of representative SEDs and median polarization properties does not fully capture the diversity of individual sources or their temporal variability. Consequently, the predicted fluxes, PD values, and duty cycles should be regarded as indicative estimates rather than precise forecasts of individual source detectability. Nevertheless, this approach provides a consistent framework for comparing a large sample of blazars and identifying promising candidates for future observations, whose detectability will ultimately depend on their actual spectral and polarimetric properties at the time of observation.

The REDSoX and GOSoX energy range opens a new observational window for probing the physical conditions and particle content of relativistic jets, extending X-ray polarimetric observations below $1 \ \rm keV$ for the first time.
This study highlights the importance of comparing simultaneous polarimetric measurements in different energy bands, in particular within the X-ray range.
Thanks to these new instruments, soft X-ray polarimetry has the potential to become a key tool in advancing our understanding of relativistic jet physics in the forthcoming decade.

\section*{Data availability}

Supplementary tables reporting the whole sample obtained can be found on the \href{https://zenodo.org/records/23057566}{Zenodo} platform.

\begin{acknowledgements}

      This research has made use of data from the RoboPol programme, a collaboration between Caltech, the University of Crete, IA-FORTH, IUCAA, the MPIfR, and the Nicolaus Copernicus University, which was conducted at Skinakas Observatory in Crete, Greece.

      Part of this work is based on archival data, software or online services provided by the Space Science Data Center - ASI.
      
      Some of the data used in this work were obtained from the MMDC.

      S. C. and I. L. were funded by the European Union ERC-2022-STG - BOOTES - 101076343. Views and opinions expressed are however those of the author(s) only and do not necessarily reflect those of the European Union or the European Research Council Executive Agency. Neither the European Union nor the granting authority can be held responsible for them.

      H. Z. is supported by NASA under award number 80GSFC24M0006 and Fermi GI program Cycle 18, grant No. 80NSSC26K0168.

      This research was partially funded by the Deutsche Forschungsgemeinschaft (DFG, German Research Foundation) as part of the DFG Research Unit FOR5195 – project number 443220636.

\end{acknowledgements}

\bibliographystyle{aa}
\bibliography{bibliography.bib}

\newpage

\begin{appendix}
\onecolumn

\section{Polarization models}
\label{appendix_polarisation_models}

We report here the polarization models used for ISP and HSP sources, obtained from BZB~J0211+1051 and Mrk~501.

\begin{figure}[h!]
    \centering
    \begin{subfigure}{0.49\linewidth}
        \centering
        \includegraphics[width=\linewidth]{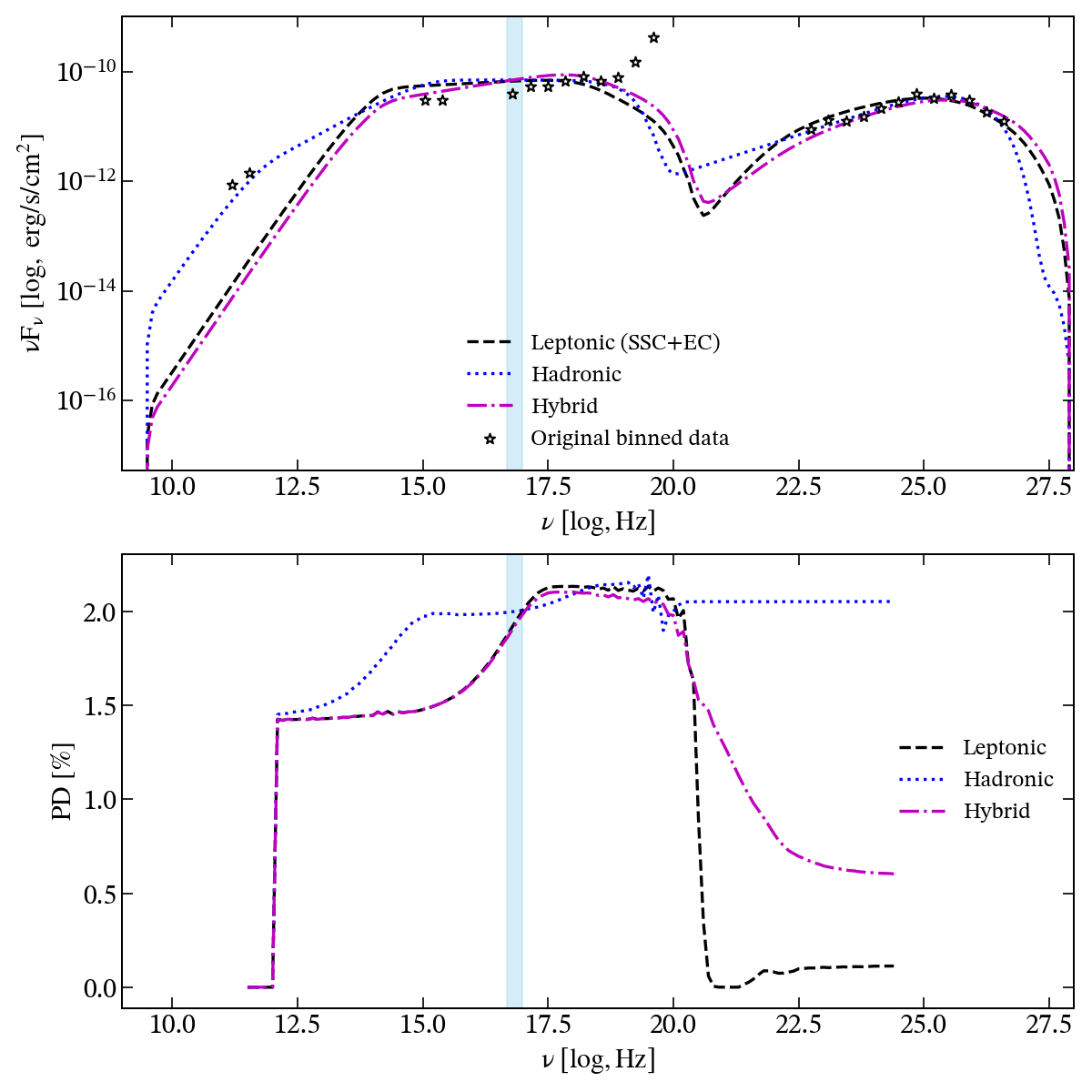}
        \caption{Mrk~501 (HSP)}
        \label{fig:mrk501_models}
    \end{subfigure}
    \hfill
    \begin{subfigure}{0.49\linewidth}
        \centering
        \includegraphics[width=\linewidth]{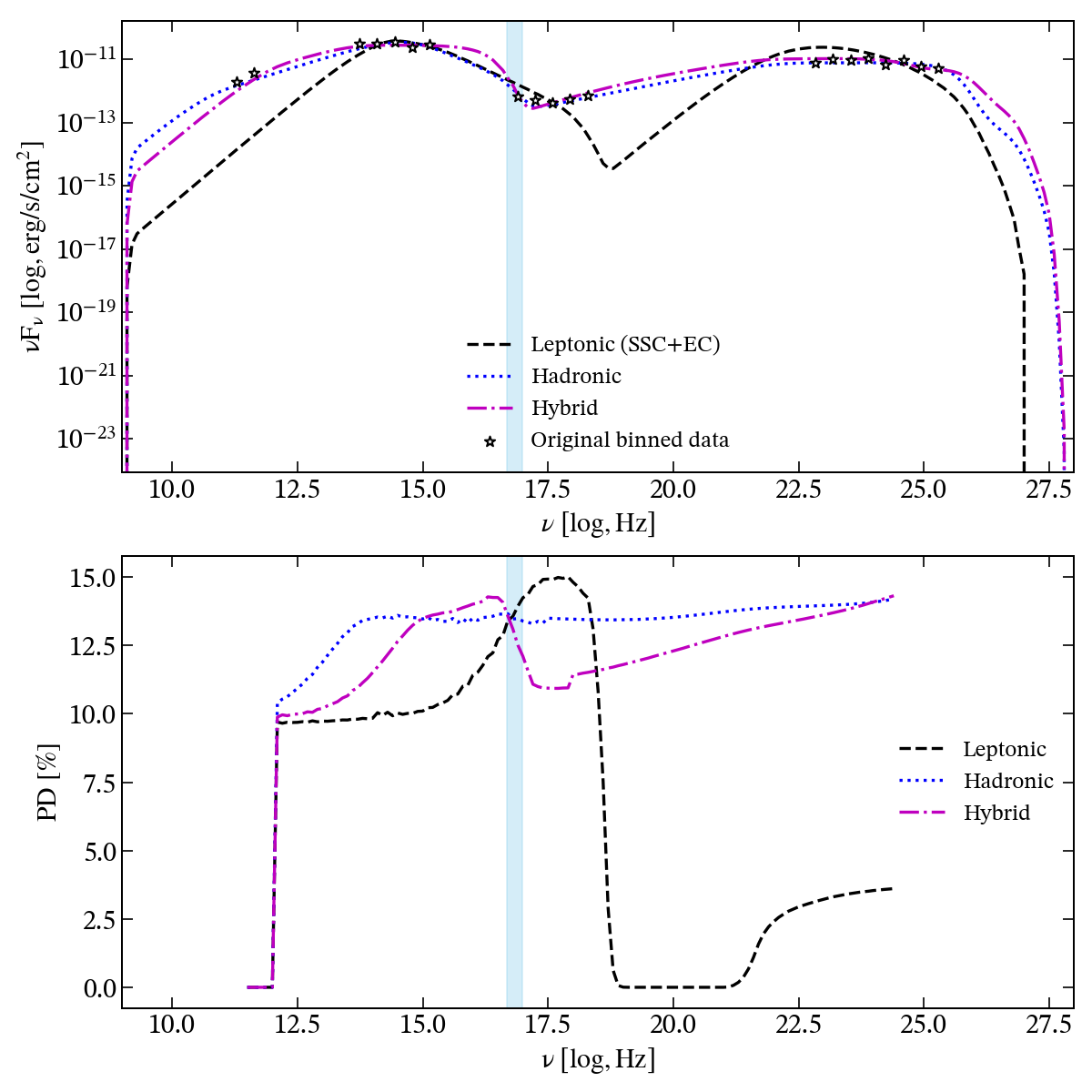}
        \caption{BZB~J0211+1051 (ISP)}
        \label{fig:bzb_models}
    \end{subfigure}

    \caption{
    Reference models used in this work (HSP on the left, ISP on the right).
    The models for LSP can be found in the text (Fig. \ref{fig:bllac_model}).
    \textit{Top panels:} Fitted SED obtained by applying the leptonic (dashed black line), hadronic (dotted blue line), and hybrid (dash-dot magenta line) models, with the black stars representing the frequency-binned data.
    \textit{Bottom panels:} Polarization degree predicted by the three models.
    }
    \label{fig:reference_models}
\end{figure}

\newpage

\section{Results for RoboPol sample}
\label{appendix_results_robopol}

We report here the results for the sources from the RoboPol Monitoring Program sample.
Sources are listed in order of decreasing detectability.
For HSP sources, we report the polarization results for all three models; for LSP and ISP sources, where the band of REDSoX and GOSoX falls into the synchrotron peak, we only report the leptonic model results.
The columns represent the source name in the Fermi 4FGL catalog, an alternative name, the flux and error in the $0.2-0.4 \ \rm keV$ band, the predicted PD and error, the MDP computed at that flux, and the duty cycle.
The complete tables are available on the \href{https://zenodo.org/records/23057566}{Zenodo} platform.

\begin{center}
\captionof{table}{Predicted REDSoX detectability for the HSP RoboPol sources.}
\label{tab:robopol_hsp}
\begin{tabular}{llccccc}
\toprule
Fermi name & Alternative name & Model & $F_{0.2-0.4\,\mathrm{keV}}$ $[\mathrm{erg\,cm^{-2}\,s^{-1}}]$ & PD [\%] & MDP [\%] & Duty cycle [\%] \\
\midrule
4FGL J1104.4+3812 & Mrk421 & Leptonic & $1.94_{-1.37}^{+1.37}\times10^{-10}$ & $8.38_{-0.00}^{+0.00}$ & $1.19$ & $100.0$ \\
4FGL J1104.4+3812 & Mrk421 & Hadronic & $1.94_{-1.37}^{+1.37}\times10^{-10}$ & $6.57_{-0.00}^{+0.00}$ & $1.19$ & $100.0$ \\
4FGL J1104.4+3812 & Mrk421 & Hybrid & $1.94_{-1.37}^{+1.37}\times10^{-10}$ & $8.20_{-0.00}^{+0.00}$ & $1.19$ & $100.0$ \\
4FGL J2000.0+6508 & 1ES1959+650 & Leptonic & $3.13_{-1.47}^{+1.47}\times10^{-10}$ & $5.98_{-0.15}^{+0.36}$ & $0.94$ & $99.0$ \\
4FGL J2000.0+6508 & 1ES1959+650 & Hadronic & $3.13_{-1.47}^{+1.47}\times10^{-10}$ & $4.48_{-0.11}^{+0.27}$ & $0.94$ & $99.0$ \\
4FGL J2000.0+6508 & 1ES1959+650 & Hybrid & $3.13_{-1.47}^{+1.47}\times10^{-10}$ & $5.89_{-0.14}^{+0.36}$ & $0.94$ & $99.0$ \\
4FGL J1838.8+4802 & GB6J1838+4802 & Leptonic & $2.51_{-2.53}^{+2.53}\times10^{-11}$ & $8.97_{-1.74}^{+1.16}$ & $3.32$ & $84.4$ \\
4FGL J1838.8+4802 & GB6J1838+4802 & Hadronic & $2.51_{-2.53}^{+2.53}\times10^{-11}$ & $6.51_{-1.26}^{+0.84}$ & $3.32$ & $84.4$ \\
4FGL J1838.8+4802 & GB6J1838+4802 & Hybrid & $2.51_{-2.53}^{+2.53}\times10^{-11}$ & $8.67_{-1.68}^{+1.12}$ & $3.32$ & $84.4$ \\
4FGL J0045.3+2128 & GB6J0045+2127 & Leptonic & $2.09_{-1.84}^{+1.84}\times10^{-11}$ & $6.05_{-0.42}^{+0.56}$ & $3.64$ & $75.1$ \\
4FGL J0045.3+2128 & GB6J0045+2127 & Hadronic & $2.09_{-1.84}^{+1.84}\times10^{-11}$ & $4.53_{-0.32}^{+0.42}$ & $3.64$ & $75.1$ \\
4FGL J0045.3+2128 & GB6J0045+2127 & Hybrid & $2.09_{-1.84}^{+1.84}\times10^{-11}$ & $5.79_{-0.40}^{+0.54}$ & $3.64$ & $75.1$ \\
4FGL J2131.5-0916 & RBS1752 & Leptonic & $3.19_{-0.31}^{+0.31}\times10^{-12}$ & $11.07_{-1.23}^{+1.66}$ & $9.31$ & $66.7$ \\
4FGL J2131.5-0916 & RBS1752 & Hadronic & $3.19_{-0.31}^{+0.31}\times10^{-12}$ & $8.18_{-0.91}^{+1.23}$ & $9.31$ & $66.7$ \\
4FGL J2131.5-0916 & RBS1752 & Hybrid & $3.19_{-0.31}^{+0.31}\times10^{-12}$ & $10.93_{-1.21}^{+1.64}$ & $9.31$ & $66.7$ \\
4FGL J1555.7+1111 & PG1553+113 & Leptonic & $1.95_{-1.24}^{+1.24}\times10^{-11}$ & $4.16_{-0.00}^{+0.00}$ & $3.76$ & $60.4$ \\
4FGL J1555.7+1111 & PG1553+113 & Hadronic & $1.95_{-1.24}^{+1.24}\times10^{-11}$ & $3.08_{-0.00}^{+0.00}$ & $3.76$ & $60.4$ \\
4FGL J1555.7+1111 & PG1553+113 & Hybrid & $1.95_{-1.24}^{+1.24}\times10^{-11}$ & $4.11_{-0.00}^{+0.00}$ & $3.76$ & $60.4$ \\
4FGL J0303.4-2407 & PKS0301-243 & Leptonic & $3.10_{-2.75}^{+2.75}\times10^{-12}$ & $9.79_{-1.22}^{+1.79}$ & $9.44$ & $50.9$ \\
4FGL J0303.4-2407 & PKS0301-243 & Hadronic & $3.10_{-2.75}^{+2.75}\times10^{-12}$ & $7.28_{-0.90}^{+1.33}$ & $9.44$ & $50.9$ \\
4FGL J0303.4-2407 & PKS0301-243 & Hybrid & $3.10_{-2.75}^{+2.75}\times10^{-12}$ & $9.68_{-1.20}^{+1.77}$ & $9.44$ & $50.9$ \\
4FGL J1725.0+1152 & 1H1720+117 & Leptonic & $3.45_{-2.00}^{+2.00}\times10^{-12}$ & $8.17_{-0.49}^{+0.85}$ & $8.95$ & $42.2$ \\
4FGL J1725.0+1152 & 1H1720+117 & Hybrid & $3.45_{-2.00}^{+2.00}\times10^{-12}$ & $8.13_{-0.49}^{+0.85}$ & $8.95$ & $42.2$ \\
\bottomrule
\end{tabular}
\end{center}

\begin{center}
\captionof{table}{Predicted REDSoX detectability for the ISP RoboPol sources.}
\label{tab:robopol_isp}
\begin{tabular}{llcccc}
\toprule
Fermi name & Alternative name & $F_{0.2-0.4\,\mathrm{keV}}$ $[\mathrm{erg\,cm^{-2}\,s^{-1}}]$ & PD [\%] & MDP [\%] & Duty cycle [\%] \\
\midrule
4FGL J2150.1-1410 & TXS2147-144 & $5.94_{-2.31}^{+2.31}\times10^{-11}$ & $3.02_{-0.12}^{+0.12}$ & $1.18$ & $94.9$ \\
4FGL J1748.6+7005 & S41749+70 & $1.32_{-0.43}^{+0.43}\times10^{-12}$ & $12.76_{-0.88}^{+1.33}$ & $7.93$ & $80.3$ \\
4FGL J0211.2+1051 & BZBJ0211+1051 & $7.00_{-5.25}^{+5.25}\times10^{-13}$ & $13.78_{-0.91}^{+2.13}$ & $10.88$ & $63.5$ \\
4FGL J1058.6+5627 & TXS1055+567 & $4.98_{-5.83}^{+5.83}\times10^{-12}$ & $5.17_{-1.44}^{+0.79}$ & $4.08$ & $57.5$ \\
4FGL J1248.3+5820 & PG1246+586 & $3.23_{-0.92}^{+0.92}\times10^{-12}$ & $4.74_{-0.71}^{+0.71}$ & $5.07$ & $42.2$ \\
\bottomrule
\end{tabular}
\end{center}

\begin{center}
\captionof{table}{Predicted REDSoX detectability for the LSP RoboPol sources.}
\label{tab:robopol_lsp}
\begin{tabular}{llcccc}
\toprule
Fermi name & Alternative name & $F_{0.2-0.4\,\mathrm{keV}}$ $[\mathrm{erg\,cm^{-2}\,s^{-1}}]$ & PD [\%] & MDP [\%] & Duty cycle [\%] \\
\midrule
4FGL J2202.7+4216 & BLLAC & $8.70_{-6.61}^{+6.61}\times10^{-12}$ & $7.18_{-0.67}^{+0.49}$ & $3.09$ & $89.2$ \\
4FGL J0137.0+4751 & OC457 & $1.05_{-0.29}^{+0.29}\times10^{-12}$ & $7.89_{-1.10}^{+1.05}$ & $8.90$ & $41.7$ \\
4FGL J1558.8+5625 & TXS1557+565 & $1.15_{-0.93}^{+0.93}\times10^{-12}$ & $6.42_{-0.50}^{+0.50}$ & $8.49$ & $29.5$ \\
\bottomrule
\end{tabular}
\end{center}

\newpage

\section{Results for the CAZ-BlazEr1 match}
\label{appendix_results_caz_erosita}

We report here the results for the sources obtained from the CAZ--BlazEr1 catalog match.
The sources were selected as explained in Sect. \ref{sect_sources_caz_erosita}.

\begin{figure}[h!]
    \centering
    \includegraphics[width=0.99\linewidth]{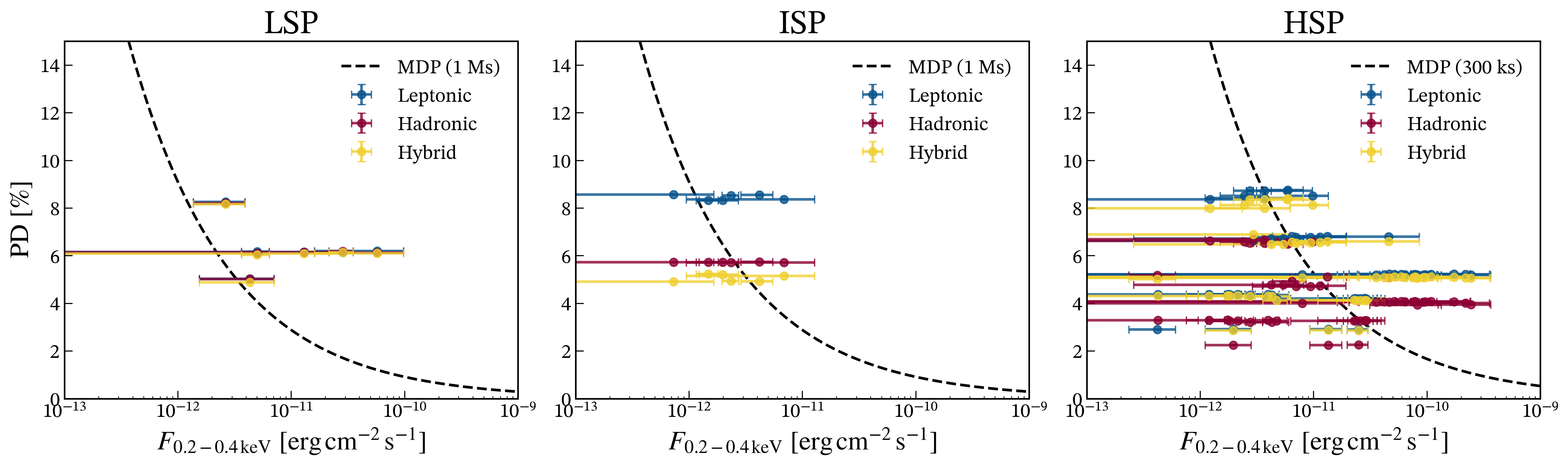}
    \caption{Red dots represent the hadronic model, blue dots the leptonic model, and yellow dots the hybrid model.
    The black dashed curve represents the MDP calculated for a given exposure time (300 ks for HSP, 1 Ms for LSP and ISP).
    }
    \label{fig:erosita_results}
\end{figure}

\begin{figure}[h!]
    \centering
    \includegraphics[width=0.5\linewidth]{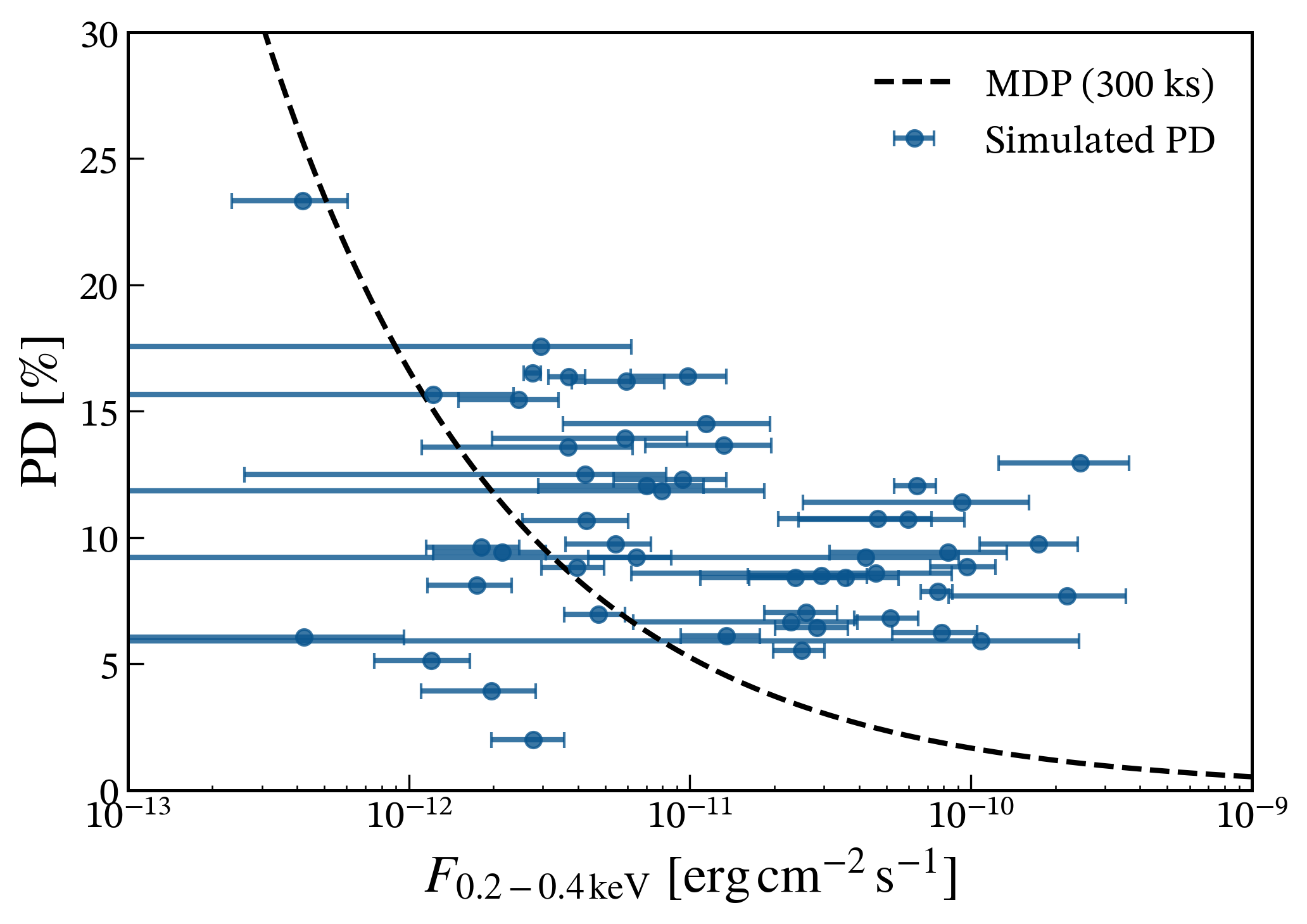}
    \caption{Predictions on HSP obtained by multiplying the assigned optical PD by 2.5 (median of the ratio of X-ray-to-optical PD ratio).
    The exposure time used to calculate the MDP (dashed black line) is 300 ks.
    }
    \label{fig:erosita_simulated_results}
\end{figure}

Fig. \ref{fig:erosita_results} represents the results obtained from shifting the leptonic, hadronic and hybrid polarization models.
Fig. \ref{fig:erosita_simulated_results} represents the HSP results obtained with the alternative method applied, i.e., multplying the assigned optical PD value by the median value of the X-ray-to-optical ratio obtained in \cite{cape2025}.

We report here the results for the sources obtained from the CAZ-BlazEr1 catalog match.
Sources are listed in order of decreasing detectability.
For HSP sources, we report the polarization results for all three models; for LSP and ISP sources, where the band of REDSoX and GOSoX falls into the synchrotron peak, we only report the leptonic model results.
The columns represent the source name in the Fermi 4FGL catalog, the flux and error in the $0.2-0.4 \ \rm keV$ band, the predicted PD, the MDP computed at that flux, and the duty cycle.
We report, for every table, only the first 10 lines.
The complete tables are available on the \href{https://zenodo.org/records/23057566}{Zenodo} platform.

\begin{center}
\captionof{table}{Predicted REDSoX detectability for the HSP CAZ-eROSITA sources.}
\label{tab:caz_hsp}
\begin{tabular}{lccccc}
\toprule
Fermi name & Model & $F_{0.2-0.4\,\mathrm{keV}}$ $[\mathrm{erg\,cm^{-2}\,s^{-1}}]$ & PD [\%] & MDP [\%] & Duty cycle [\%] \\
\midrule
4FGLJ1057.8-2754 & Leptonic & $3.70_{-0.55}^{+0.55}\times10^{-12}$ & $8.72$ & $8.65$ & $100.0$ \\
4FGLJ1057.8-2754 & Hybrid & $3.70_{-0.55}^{+0.55}\times10^{-12}$ & $8.34$ & $8.65$ & $100.0$ \\
4FGLJ0558.0-3837 & Leptonic & $9.70_{-2.53}^{+2.53}\times10^{-11}$ & $5.19$ & $1.69$ & $99.6$ \\
4FGLJ0558.0-3837 & Hybrid & $9.70_{-2.53}^{+2.53}\times10^{-11}$ & $5.08$ & $1.69$ & $99.6$ \\
4FGLJ1823.6-3453 & Leptonic & $1.74_{-0.66}^{+0.66}\times10^{-10}$ & $5.22$ & $1.26$ & $99.5$ \\
4FGLJ0650.7+2503 & Leptonic & $2.46_{-1.20}^{+1.20}\times10^{-10}$ & $5.19$ & $1.06$ & $99.4$ \\
4FGLJ0650.7+2503 & Hybrid & $2.46_{-1.20}^{+1.20}\times10^{-10}$ & $5.05$ & $1.06$ & $99.4$ \\
4FGLJ1823.6-3453 & Hybrid & $1.74_{-0.66}^{+0.66}\times10^{-10}$ & $5.09$ & $1.26$ & $99.4$ \\
4FGLJ0449.4-4350 & Leptonic & $9.82_{-3.66}^{+3.66}\times10^{-12}$ & $8.50$ & $5.30$ & $99.3$ \\
4FGLJ1823.6-3453 & Hadronic & $1.74_{-0.66}^{+0.66}\times10^{-10}$ & $4.07$ & $1.26$ & $99.3$ \\
\bottomrule
\end{tabular}
\end{center}

\begin{center}
\captionof{table}{Predicted REDSoX detectability for the ISP CAZ-eROSITA sources.}
\label{tab:caz_isp}
\begin{tabular}{lcccc}
\toprule
Fermi name & $F_{0.2-0.4\,\mathrm{keV}}$ $[\mathrm{erg\,cm^{-2}\,s^{-1}}]$ & PD [\%] & MDP [\%] & Duty cycle [\%] \\
\midrule
4FGLJ0847.2+1134 & $4.19_{-1.32}^{+1.32}\times10^{-12}$ & $8.55$ & $4.45$ & $96.9$ \\
4FGLJ1213.3-2618 & $6.89_{-5.94}^{+5.94}\times10^{-12}$ & $8.36$ & $3.47$ & $93.1$ \\
4FGLJ0414.8-5338 & $2.35_{-0.39}^{+0.39}\times10^{-12}$ & $8.52$ & $5.94$ & $89.4$ \\
4FGLJ0955.1+3551 & $1.98_{-0.75}^{+0.75}\times10^{-12}$ & $8.33$ & $6.47$ & $73.0$ \\
4FGLJ0352.9-3623 & $1.49_{-0.33}^{+0.33}\times10^{-12}$ & $8.31$ & $7.47$ & $63.3$ \\
4FGLJ2325.4-4800 & $7.30_{-9.28}^{+9.28}\times10^{-13}$ & $8.56$ & $10.66$ & $39.8$ \\
\bottomrule
\end{tabular}
\end{center}

\begin{center}
\captionof{table}{Predicted REDSoX detectability for the LSP CAZ-eROSITA sources.}
\label{tab:caz_lsp}
\begin{tabular}{lcccc}
\toprule
Fermi name & $F_{0.2-0.4\,\mathrm{keV}}$ $[\mathrm{erg\,cm^{-2}\,s^{-1}}]$ & PD [\%] & MDP [\%] & Duty cycle [\%] \\
\midrule
4FGLJ0433.0+0522 & $2.84_{-0.69}^{+0.69}\times10^{-11}$ & $6.14$ & $1.71$ & $99.9$ \\
4FGLJ1229.0+0202 & $5.73_{-4.12}^{+4.12}\times10^{-11}$ & $6.19$ & $1.20$ & $99.0$ \\
4FGLJ0522.9-3628 & $1.30_{-1.63}^{+1.63}\times10^{-11}$ & $6.16$ & $2.52$ & $93.9$ \\
4FGLJ2156.0-6942 & $5.03_{-1.37}^{+1.37}\times10^{-12}$ & $6.16$ & $4.06$ & $93.9$ \\
4FGLJ0854.8+2006 & $2.65_{-1.27}^{+1.27}\times10^{-12}$ & $8.26$ & $5.60$ & $85.7$ \\
4FGLJ1256.1-0547 & $4.32_{-2.77}^{+2.77}\times10^{-12}$ & $5.03$ & $4.38$ & $58.5$ \\
\bottomrule
\end{tabular}
\end{center}

We report here the results for the HSP CAZ and eROSITA cross-match, obtained with the alternative method explained in Sect. \ref{sect_sources_caz_erosita}.
These results do not depend on the polarization model considered.
Sources are listed in order of decreasing detectability.
We report only the first 10 sources.
The complete table is available on the \href{https://zenodo.org/records/23057566}{Zenodo} platform.

\begin{center}
\captionof{table}{Predicted REDSoX detectability for the HSP CAZ-eROSITA sources.}
\label{tab:caz_hsp_alternative}
\begin{tabular}{lcccc}
\toprule
Fermi name & $F_{0.2-0.4\,\mathrm{keV}}$ $[\mathrm{erg\,cm^{-2}\,s^{-1}}]$ & PD [\%] & MDP [\%] & Duty cycle [\%] \\
\midrule
4FGLJ1057.8-2754 & $3.70_{-0.55}^{+0.55}\times10^{-12}$ & $16.36$ & $8.65$ & $99.8$ \\
4FGLJ0650.7+2503 & $2.46_{-1.20}^{+1.20}\times10^{-10}$ & $12.94$ & $1.06$ & $99.4$ \\
4FGLJ1221.3+3010 & $2.20_{-1.37}^{+1.37}\times10^{-10}$ & $7.69$ & $1.12$ & $99.4$ \\
4FGLJ0449.4-4350 & $9.82_{-3.66}^{+3.66}\times10^{-12}$ & $16.37$ & $5.30$ & $99.3$ \\
4FGLJ1823.6-3453 & $1.74_{-0.66}^{+0.66}\times10^{-10}$ & $9.75$ & $1.26$ & $99.3$ \\
4FGLJ0143.7-5846 & $6.42_{-1.09}^{+1.09}\times10^{-11}$ & $12.03$ & $2.08$ & $99.2$ \\
4FGLJ0558.0-3837 & $9.70_{-2.53}^{+2.53}\times10^{-11}$ & $8.84$ & $1.69$ & $99.1$ \\
4FGLJ0209.3-5228 & $4.59_{-3.97}^{+3.97}\times10^{-11}$ & $8.58$ & $2.45$ & $98.9$ \\
4FGLJ1010.2-3119 & $7.61_{-0.97}^{+0.97}\times10^{-11}$ & $7.86$ & $1.91$ & $98.8$ \\
4FGLJ0416.9+0105 & $7.88_{-2.62}^{+2.62}\times10^{-11}$ & $6.23$ & $1.87$ & $98.7$ \\
\bottomrule
\end{tabular}
\end{center}

\end{appendix}

\end{document}